\documentclass[
    ,final            % use final for the camera ready runs
  ]
  {aipproc}

\layoutstyle{6x9}
\def \lta {\mathrel{\vcenter
     {\hbox{$<$}\nointerlineskip\hbox{$\sim$}}}}

\begin{document}

\title{Theories with Extra Dimensions\footnote{Presented at The 15th
International Topical Conference on  
Hadron Collider Physics, East Lansing, Michigan, June 14-18 2004. }}

\author{Gustavo Burdman\footnote{Email: burdman@if.usp.br}}{
  address={Theory Group, Lawrence Berkeley National Laboratory, Berkeley, 
CA 94720\footnote{Current address: Instituto de F\'isica, 
Universidade de S\~ao Paulo, S\~ao Paulo, Brazil. }}
}

\begin{abstract}
We review some aspects of theories with compact extra dimensions. 
We consider the motivation and the theoretical basis of Large, Universal and Warped Extra 
Dimensions.
We focus on those aspects that are potentially relevant  
in the phenomenology at colliders.
\end{abstract}

\maketitle

%%%%%%%%%%%%%%%%%%%%%%%%%%%%%%%%%%%%%%%%%%%%
%% MAINMATTER
%%%%%%%%%%%%%%%%%%%%%%%%%%%%%%%%%%%%%%%%%%%%

\section{The Need for Physics Beyond the Standard Model}
Although the Standard Model (SM) is in quite good agreement with experiment so far, 
there are good reasons to believe its validity is limited to energies perhaps as low as
1 TeV. As far as the SM goes, all we need is the Higgs boson. Its appearance below the 
TeV scale is required by unitary. The discovery of just a Higgs boson at the TeV scale, although 
confirming the SM, would be rather puzzling. This is because our understanding of quantum field 
theory tells us that this is a very unnatural and fine-tuned scenario. 
\begin{figure}[h]
  \includegraphics[height=1in]{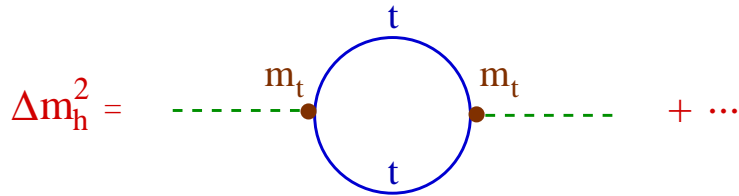}
  \caption{Top loop contribution to the Higgs mass.}
\label{toploop}
\end{figure}
To see this let us consider radiative corrections to the Higgs mass. These come, at leading 
order, from loops containing fermions (most notably the top quark, see Figure~\ref{toploop}), 
the gauge bosons, and the 
Higgs itself. To this order, $\Delta m_h^2$ is given by 
\begin{equation}
{\Delta m_h^2} \simeq \,{\Lambda^2}\,\frac{{3}\left(2{ M_W^2}+{ M_Z^2}
+{ m_h^2}-4{ m_t^2}\right)}{({32\pi^2v^2})},
\label{mh2}
\end{equation}
plus logarithmically sensitive terms. 
Thus, the Higgs mass squared is quadratically sensitive to the ultra-violet cutoff of the 
SM. If this were to be a theory valid all the way to $M_P$, then either the natural scale for 
$m_h$ is the Planck mass $M_P$ or a rather fine-tuned cancellation must take place  
between the bare Higgs mass and the sum of all corrections to all orders. 
This cancellation -of about 1 in $10^{17}$- is why we refer to the SM as fine-tuned. 
If this does not take place, then why is the weak scale so much smaller 
than the Planck scale ? This is the hierarchy problem of the SM.
 
If physics beyond the SM is to solve the hierarchy problem, it has to come not far above the 
TeV scale, so that the corrections of eqn.(\ref{mh2}) do not result in  $m_h> 1~$TeV. 
Classic examples of solutions to the hierarchy problem are theories with supersymmetry  at the 
weak scale~\cite{susy} and Technicolor theories~\cite{techni}. More recently, the idea that 
compact extra spatial dimensions, with typical compactification radii considerably larger than 
the Planck length, could be the solution to the hierarchy problem has gained a lot of attention.
In what follows, we review three scenarios where this is the case: 
the possibility of the existence of  Large Extra Dimensions (LED)~\cite{led1}, Universal 
Extra Dimensions (UED)~\cite{ued1} and finally the so-called Warped Extra Dimensions 
(WED)~\cite{wed1}. We emphasize the theoretical aspects of these scenarios which have some 
bearing on their experimental signals at hadron colliders.

\section{Large Extra Dimensions}
A rather innovative solution to the hierarchy problem was proposed in Ref.~\cite{led1}: 
gravity appears weak because it propagates in the volume of compact extra dimensions, 
not available to the rest of the SM fields. The extra volume dilutes gravity's strength. 
The Planck scale is not a fundamental parameter in the extra-dimensional Einstein's action, 
but a scale derived from a volume suppression. The fundamental scale of gravity is $M_*$, 
and it satisfies 
\begin{equation}
M_P^2 = M_*^{n+2} \, R^n, 
\label{gauss}
\end{equation}
where $n$ is the number of extra dimensions, and $R$ is the average compactification radius. 
Then, in principle, the fundamental scale of gravity could be much smaller than $M_P$ 
if the extra dimensions are big enough. For instance, if we have in mind solving the hierarchy 
problem, then we could choose $M_*\simeq 1$~TeV. Then, 
\begin{equation}
R \sim 2\cdot 10^{-17}\,10^{\frac{32}{n}}{\rm cm}.
\label{rincm}
\end{equation}
Thus, if we take $n=1$, we get that $R\sim10^8~$Km, which is certainly excluded. 
Taking $n=2$, one has $R\sim 1$~mm, which is a distance scale 
already constrained by Cavendish-type experiments. For $n>2$, we need $R<10^{-6}$~mm, 
which is not going to be reached by gravity experiments any time soon.

\begin{figure}[h]
  \includegraphics[height=0.8in]{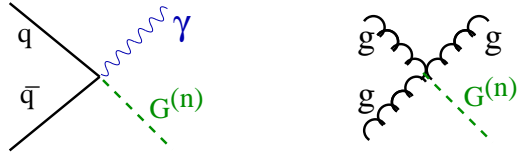}
  \caption{Production of KK Gravitons.}
\label{mono}
\end{figure}

The fact that gravity propagates in compact extra dimensions leads to the existence of 
graviton excitations with a mass gap given by $\Delta m\sim 1/R$. Then, in this scenario, 
there are new states, with spin 2, and with rather small masses. For instance, for $n=2$
the Kaluza-Klein graviton mass starts at about $10^{-3}$~eV, and for $n=3$ at about 
$100$~eV. Then, although the couplings of graviton excitations to matter are 
gravitationally suppressed, these states are so copiously produced at high energies 
($E\gg 1/R$) that when we sum over all these final states, the inclusive cross
sections are not suppressed by $M_P$, but by $M_*$:
\begin{equation} 
\sigma \sim \frac{E^{n}}{M_*^{{n}+2}}. 
\end{equation}
On the other hand, since KK graviton lifetimes are still $M_P^2$ suppressed they would 
escape detection, leaving large missing energy signals as their mark. The processes that most 
uniquely would point to this physics at hadron colliders are of the mono-jet type, as depicted 
in Figure~\ref{mono}.
Limits from the Tevatron already are available~\cite{karagoz} and are already around 
$O(1)$~TeV, depending on the number of dimensions. 
Another signal of LED comes from the virtual exchange of KK gravitons. 
This induces dim-6 operators of the form 
$(\bar q\gamma_\mu\gamma_5 q)(\bar f\gamma_\mu\gamma_5 f)$ 
entering, for example, in Drell-Yan production. 
Moreover, dim-8 operators ($T_{\mu\nu} T^{\mu\nu}$), result in 
$\bar ff\to\gamma\gamma, ZZ,\cdots$.
For $n=2$, the bounds on $M_*$ from the contributions of these operators are in the multi-TeV
region already~\cite{karagoz}. 

Astrophysical constraints have played an important role in the viability of the LED scenario.
The most tight bound comes from Supernova cooling, where graviton KK emission could cool 
the supernova too fast. For instance, for $n=2$ this requires $M_*>(10-100)$~TeV.

Finally, the proposal that the fundamental scale  of gravity, $M_*$, might not be far above
the TeV scale, raises the possibility that strong gravity effects -such as black holes- 
might be visible at colliders experiments~\cite{bhled}, 
or in ultra high energy cosmic rays~\cite{crled}.

\section{Universal Extra Dimensions}
If  - unlike in the LED scenario - we allow fields other than the graviton to propagate in the 
extra dimensions, then constraints on $1/R$ are much more severe. This is because the couplings
of all other fields are not suppressed gravitationally, but at most by the weak scale. 
Naively, we would expect then that bounds on $1/R$ climb rapidly to $O(1)$~TeV, and this is what 
happens~\cite{karagoz} if {\em some} of the SM fields other than the graviton are in the bulk. 

\begin{figure}[h]
  \includegraphics[height=0.8in]{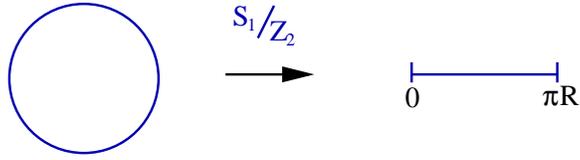}
  \caption{Orbifold Compactification.}
\label{orbi}
\end{figure}

However, it was shown in Ref.~\cite{ued1} that if {\em all } fields propagate in the extra 
dimensional bulk (universal extra dimensions), then bounds on $1/R$ drop to 
considerably lower values. The reason is that -upon compactification- momentum conservation leads
to KK-number conservation. For instance, in 5D, $p_5$ the fifth component of the 5d momentum
is quantified and given by $p_5=n/R$, with $n$ the KK number. Among other things, this implies 
that in UED, KK modes cannot be singly produced, but they must be pair produced. 
This raises the bounds both from direct searches, as well as those from electroweak precision 
constraints~\cite{ued1}. Furthermore, compactification must be realized on an orbifold in 
order to allow for chiral fermion zero-modes, such as the ones we observe in the SM.
In 5D, for example, this means an $S_1/Z_2$ compactification as illustrated in Figure~\ref{orbi}.
As a consequence of the orbifolding, KK-number conservation is broken, but there is a 
remnant left:  the parity of the KK modes must be conserved at the vertices. This 
KK-parity is very similar to R-parity in supersymmetric theories. It still means that 
zero-modes cannot be combined to produce a single KK excitation (for instance in an s-channel). 
In addition, KK-parity conservation means that the lightest state of the 
first KK level cannot decay into zero-modes, making it stable and a candidate for dark matter. 

Direct constraints from the Tevatron in Run I as well as electroweak precision constraints on 
oblique parameters, give 
\begin{equation}
1/R \geq \left\{\begin{array}{cc}
300{\rm ~GeV}&{\rm for ~5D} \\
(400-600){\rm ~GeV}  &{\rm for ~6D}
\end{array}
\right.
\nonumber
\end{equation}
These rather loose bounds imply that in principle even the Tevatron in Run II still has a chance 
of seeing UED. However, the signals are in general subtle. The reason is that -at leading order- 
all states in the same KK level are degenerate. Radiative corrections generate mass 
splittings~\cite{uedrad}, but these are still small enough for the energy release to be small
in the production and subsequent decay of KK states.  
For instance, if a pair of level 1 quarks is produced, each of them might decay as 
$Q_{1L}\to W^{\pm}_1Q'_L$, where the typical splitting between the $Q_{1L}$ and the $W^{\pm}_1$
might be just a few tens of GeV, depending on the number of extra dimensions (as well as the 
values of the brane kinetic terms). Thus, the quark jet tends to be soft, and this is repeated down 
the decay chain.  In Ref.~\cite{fooled}, the golden mode is identified as being 
$q\bar q\to Q_1Q_1\to Z_1Z_1 + \not E_T\to 4\ell + \not E_T$, where a large fraction of the 
missing energy is taken away by the lightest KK particle (LKP), in analogy with typical MSSM signals. 

In fact, the similarity with the MSSM makes distinguishing it from UED a challenging proposition. 
The most  distinct aspect of the UED scenario is the existence of KK levels, so the
observation of the second KK level might be what is needed to tell it apart from other 
new physics~\cite{fooled}. 
In the 5D case, for instance, level 2 KK states can decay via KK-number conserving interactions 
to either another lighter level 2 state 
plus a zero-mode, or to two level 1 states. In both cases, the energy release is very small. 
On the other hand, localized kinetic terms induce not only additional mass splitting, but also 
KK-number violating (but KK-parity conserving) interactions. These are volume suppressed, and 
therefore there will only be able to compete so much with the phase-space suppressed 
terms mentioned above. 

In the 6D case, however, the decay channel of the second KK level to two level 1 states is not 
present. This is because the generic mass of the second KK level is $M_2 = \sqrt{2}/R$, which is 
smaller than $2/R$, the sum of the masses of two level-1 states. 
Thus, in the 6D scenario, level 2 KK states 
can only decay through the localized kinetic terms and into two zero-modes.  This signal then 
may be used to distinguish the 5D and 6D cases. The energy release is much larger 
(e.g. two fast leptons of hard jets), and the level-2 KK states could be produced in the 
s-channel (00-2 preserves KK parity) through suppressed kinetic terms, which results in a 
narrow resonance~\cite{bdp}.
Then, the 6D scenario, well-motivated in its own merits (e.g. proton stability~\cite{uedsd})
could be  distinguished form the more typical MSSM-like 5D case.

\section{Warped Extra Dimensions}
\begin{figure}[h]
  \includegraphics[height=1.8in]{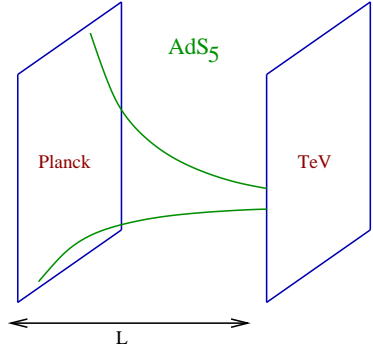}
  \caption{Warped Extra Dimension. The local cutoff is exponentially smaller than $M_P$.} 
\label{warped}
\end{figure}

\begin{figure}[h]
  \includegraphics[height=2.2in,angle=90]{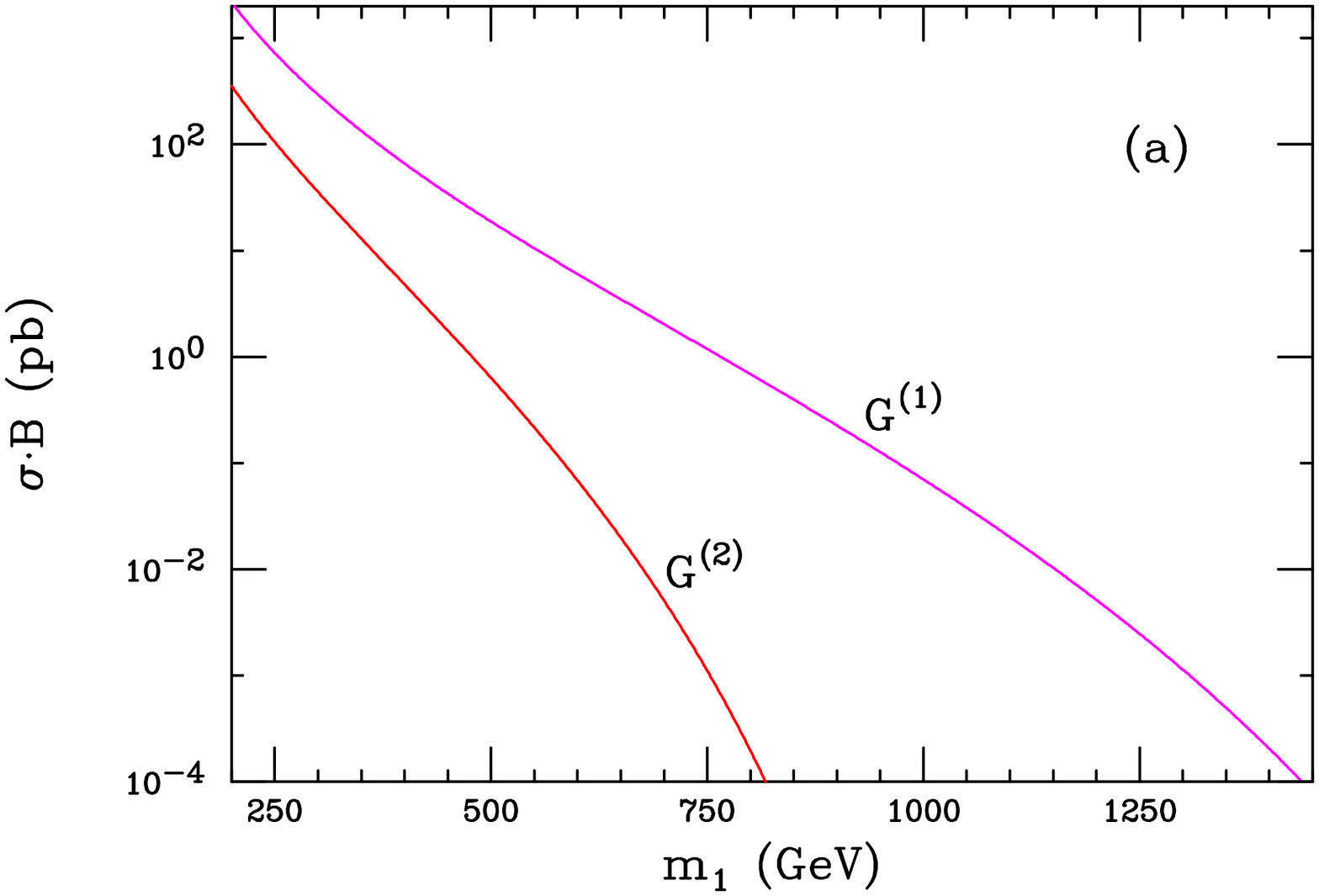}
  \hskip0.4in
  \includegraphics[height=2.2in,angle=90]{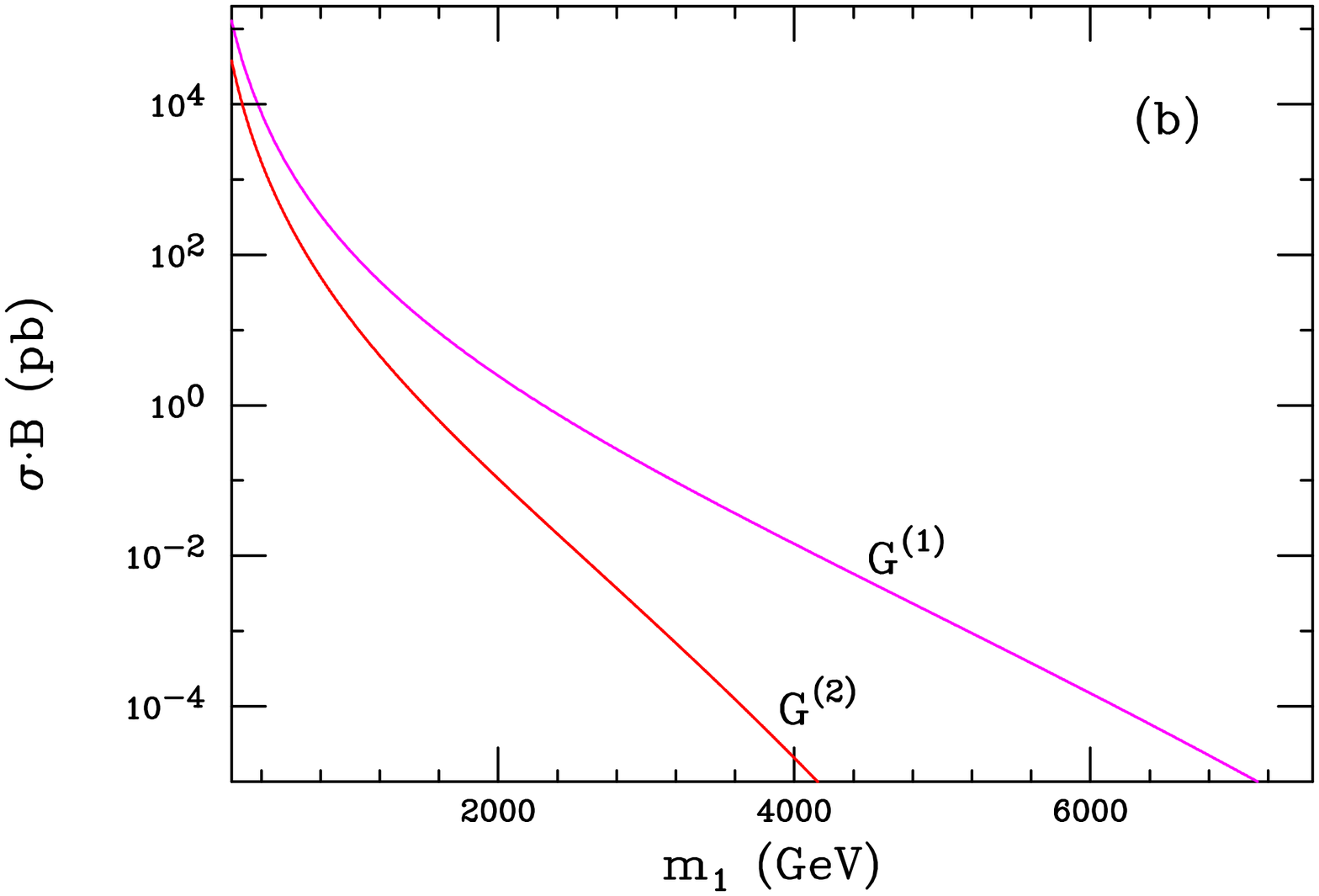}
  \caption{$\sigma\cdot Br$ for the s-channel production of the first two KK gravitons at 
the Tevatron (a), and the LHC (b). From Ref.~\cite{dhr}.} 
\label{dywed}
\end{figure}

A new solution to the hierarchy problem making use of one extra dimension was proposed in 
Ref.~\cite{wed1}. Unlike the two previous cases, the extra dimension does not have a flat metric. 
In the Randall-Sundrum (RS) setup, the metric is that of Anti-de Sitter in 5D, and is given by: 
\begin{equation}
ds^2 = e^{-2{k}|y|}\,\eta^{\mu\nu}dx_\mu dx_\nu  + dy^2,
\label{metric}
\end{equation}
which is a solution of Einstein's equations in 5D, as long as the bulk cosmological constant
is adjusted to cancel the cosmological constant on the fixed points. Then, the branes 
have a flat metric, as desired. In eqn.(\ref{metric}) $k\lta M_P$ is the $AdS_5$ curvature and 
$y$ is the coordinate of the fifth dimension. The only scale in the 5D 
Einstein-Hilbert action is $M_P$. However, when at a distance $y$ from the origin of the extra dimension,
all energies are exponentially suppressed by a factor of $\exp{(-ky)}$.
Then, if all SM fields, except gravity, were confined at a distance $L$ from the origin, the 
local cutoff would not be $M_P$ but 
\begin{equation}
\Lambda_L = M_P \,e^{-kL}.
\label{cutloc} 
\end{equation}
This is depicted in Figure~\ref{warped}. The compactification is done in the $S_1/Z_2$ orbifold, 
with $L=\pi R$. 
If we want the local cutoff to be the TeV scale, therefore explaining the hierarchy, we need to 
choose
\begin{equation}
k\,R\simeq O(10),
\label{kr10}
\end{equation}
which does not constitute a very significant fine-tuning. Then, in the RS scenario, an 
exponential hierarchy is generated by the non-trivial geometry of the extra dimension. 
This scenario already has important experimental consequences. Since gravity propagates 
in the bulk, there is a tower of KK gravitons. The zero-mode graviton has a wave-function 
localized towards the Planck brane and couples to matter with its usual coupling, suppressed by 
$1/M_P^2$. The KK gravitons, on the other hand, have masses starting at O(1)~TeV, and
couple to matter on the TeV brane as $1/(TeV)^2$. Then, KK gravitons can be produced at 
accelerators with significant cross sections. For instance, the Drell-Yan process 
would receive a contribution from s-channel KK gravitons as in 
$q\bar q\to G^{(n)}\to e^+e^-$. This leads to the cross sections of Figure~\ref{dywed}. 
 
The RS proposal solves the hierarchy problem because the radiative corrections to $m_h$ are
now cutoff at the TeV scale. The SM operates in our 4D slice. But the mechanism of EWSB is still
that of the SM. Moreover, the origin of fermion masses (the other hierarchy) is completely 
unexplained, together with a number of other issues ranging from gauge coupling unification to 
dark matter. Allowing additional fields to propagate in the bulk opens up a great deal of model 
building opportunities. In general, unless supersymmetry is invoked, the Higgs must 
remain localized in the TeV brane, or it would receive large quadratically divergent 
corrections of order of $M_P$, just as in the SM. 

If gauge fields are allowed in the bulk then their KK expansion takes the form
\begin{equation}
A_\mu(x,y) = \frac{1}{\sqrt{2\pi R}}\sum_{n=0}^{\infty} {A_\mu^{(n)}(x)} 
{\chi^{(n)}(y)}~,
\label{gaugekk}
\end{equation}
where $\chi^{(n)}(y)$ is the wave-function in the extra dimension for the 
nth KK excitation of the gauge field.  In the 4D effective theory, there is - in general - 
a zero-mode $A_\mu(x)^{(0)}$, and a KK tower of states with masses 
\begin{equation}
m_n \simeq (n -O(1))\times\pi{k} e^{-{k}\pi R},
\label{mn}
\end{equation}
starting at $O(1)$~TeV. Their wave-functions are localized towards the TeV brane. 
The gauge symmetry in the bulk must be enlarged with respect to the SM in 
order to contain an $SU(2)_L\times SU(2)_R$ symmetry. The extra $SU(2)_R$ restores 
a gauge version of custodial symmetry in the bulk, thus avoiding severe $T$ parameter 
constraints~\cite{adms} (in the dual language of the CFT, there is a global symmetry 
associated with it, the custodial symmetry). 

Just like the gauge fields, if fermions are allowed to propagate in the bulk they will
have a similar KK decomposition, given by
\begin{equation}
\Psi_{L,R}(x,y) = \frac{1}{\sqrt{2\pi R}}\,\sum_{n=0}\,{ \psi_n^{L,R}(x)} e^{2k\,|y|} 
{f_n^{L,R}(y)},
\label{fermions}
\end{equation}
where $f_n^{L,R}(y)$ are the wave-functions of the KK fermions in the extra dimension, and 
the superscripts $L$ and $R$ indicate the chirality of the KK fermion. Since fermions are 
not chiral in 5D, half of their components are projected out in the orbifold 
compactification. 
Unlike gauge fields in the bulk, fermions are allowed to have a mass term since  
there is no chiral symmetry protecting it. Then the typical, bulk fermion mass term looks like
\begin{equation}
S_f = \int d^4x~dy~ \sqrt{-g} \left\{ \cdots- ~{c}{~k} 
\bar\Psi(x,y)\Psi(x,y)\right\}~,
\end{equation}
where naturally $c\sim O(1)$, i.e. the bulk fermion mass is of the order of the 
AdS curvature scale $k$. The KK fermion wave-functions in this case have the form
\begin{equation}
{f_0^{R,L}(y)} = 
\sqrt{\frac{k\pi R\,(1\pm 2{c})}{e^{k\pi R(1\pm2{ c})}-1}}\;
e^{\pm{ c} \,k\,y}. 
\label{ferwf}
\end{equation}
Then, the localization of the KK fermion wave function in the extra dimension 
is controlled by the $O(1)$ parameter $c$ with {\em exponential} sensitivity~\cite{gp}. 
All that is needed to explain the wildly varying fermion spectrum is $O(1)$ flavor breaking 
in the bulk, which could be naturally originated at the cutoff. 
Fermions with wave-functions towards the TeV brane ($c<1/2$) will have a larger overlap with the 
Higgs VEV, and therefore a larger mass, of $O(v)$. Light fermions, on the other hand, will have 
wave-functions localized towards the Planck brane ($c>1/2)$. For $c=1/2$ the fermion 
wave-function is flat.
\begin{figure}[h]
  \includegraphics[height=1.8in]{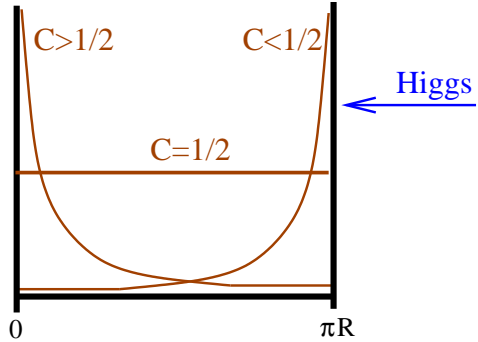}
   \caption{Fermions are localized according to the choice of the $O(1)$ parameter 
$c$, the bulk fermion mass in units of $k$, the AdS curvature.
}
\label{ferloc}
\end{figure}
This is shown in Figure~\ref{ferloc}. The need to generate a large enough value for $m_t$, 
forces us to localize the top quark not too far from the TeV brane. Even if we localize 
$t_R$ to this IR fixed point, the localization of $t_L$ -and consequently $b_L$- 
cannot be chosen to be at the Planck brane. If $b_L$ is forced to have a significant overlap 
with the TeV brane, there will be non-universal couplings of both the KK gauge bosons of all 
gauge fields, and the zero-mode weak gauge bosons ($W^\pm,Z$). 
The latter, result from the deformation of their (otherwise flat) 
wave-function  due to the Higgs VEV on the TeV brane. Fermions with profiles close to 
the TeV brane can feel this effect and couplings like $Z\to b_L\bar b_L$ 
would be affected~\cite{adms}. It would also lead to FCNCs at tree level mediated
by the $Z$, and that could be observed in $B$ decays such as $b\to\ell^+\ell^-$~\cite{bn}. 
On the other hand, the KK excitations of all gauge bosons would have non-universal couplings
to all fermions, and particularly to the top and to $b_L$. This could lead to interesting effects
in hadronic $B$ decays and $CP$ violation, especially when considering the interactions with 
KK gluons~\cite{gb}.

Finally, we could ask the question: could we get rid of the Higgs boson altogether ?
After all, it looks a bit ad hoc, localized in the TeV brane. We know that boundary conditions
(BC) can be used to break gauge symmetries in extra dimensional theories. 
It was proposed in Ref.~\cite{hless1} that the electroweak symmetry could be ``broken'' by 
BC in a 5D theory as a way to replace the Higgs field. The first question would be: what about 
the unitarity of electroweak amplitudes such as $W^+W^-$ scattering ? If the Higgs boson is not
present how are these amplitudes unitarized ? The answer is that the KK excitations of the 
gauge bosons do the job~\cite{cdh}. The actual models that (nearly) work  are similar 
to the one we had before on $AdS_5$, but without a Higgs boson on the TeV brane~\cite{hlads}. 
The BC can be thought of as obtained by the presence of brane-localized scalar fields that 
get VEVs. In the limit of these VEVs $\to\infty$, one recovers the BC. Thus although the 
origin of the BC might be a set of scalar fields getting VEVs, these need not be at low energies.
It is in this sense that the theory is Higgsless. 

\begin{figure}[h]
  \includegraphics[height=1.8in]{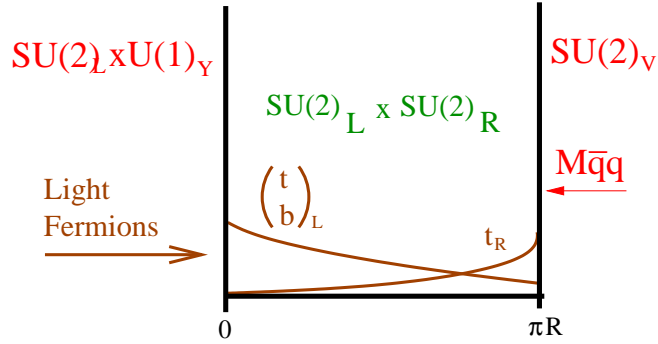}
   \caption{A Higgless model of EWSB and fermion masses. 
}
\label{higgless}
\end{figure}

The other question is how do fermions get their masses ? With the appropriate choice of 
BC the bulk gauge symmetry breaks as $SU(2)_L\times SU(2)_R\times U(1)_X \to 
U(1)_{\rm EM}$. But the BC restrict the gauge symmetry differently at different fixed points.
This can be seen in Figure~\ref{higgless}. The BC restrict the symmetry at the TeV brane to be
$SU(2)_{L+R}=SU(2)_V$, which allows us to write brane localized vector 
mass terms such as, for instance
\begin{equation}
{\cal S}_f = \int d^4x\int dy \delta(y-\pi R)\sqrt{-g}\,\left\{{M_u}q\psi_{\bar u} + 
{M_d}q\psi_{\bar d} + \cdots\right\},
\label{vecmass}
\end{equation}
where $M_u$ and $M_d$ are $\simeq O(1)$~TeV, and $q$, $\psi_{\bar u}$ and $\psi_{\bar d}$ a left-handed, 
right-handed up and right-handed down quarks respectively. Thus, the fermion mass hierarchy is 
still generated by $O(1)$ flavor breaking in the bulk fermion mass parameter $c$. The difference now is that
the masses are generated by the overlap with a vector mass term as opposed to the Higgs VEV. 
Thus from Figure~\ref{higgless} we see that the problem of flavor violation coming from the need to have 
a heavy top quark is still present here~\cite{bn}. 

\begin{figure}[h]
  \includegraphics[height=2.9in]{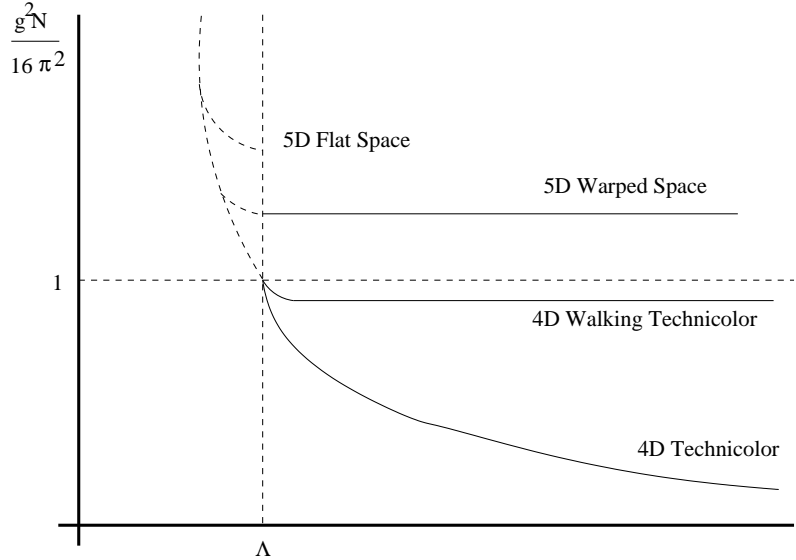}
   \caption{The 't Hooft coupling vs. energy scale for various theories. From Ref.~\cite{bn}.  
}
\label{phase}
\end{figure}

The electroweak constraints on these kinds of theories are quite important. For instance, the $S$ 
parameter is given by~\cite{bn,bpr}
\begin{equation}
{S} \sim  16\pi\frac{v^2}{m_{KK}^2} = \frac{{N}}{\pi},
\label{svsn}  
\end{equation}
where in the second equality, $N$ refers to the size of the gauge group in the dual 4D CFT. 
Thus, for large $N$, which in the $AdS_5$ side corresponds to a weakly couple KK sector, the $S$
parameter tends to be larger than experimentally acceptable. 
Several possibilities have been considered in the literature to deal with this problem. For instance, 
negative $S$ contributions might be induced by TeV localized kinetic terms~\cite{ccgt}, however 
there is always a constraining combination of $S$ and $T$~\cite{sek}.
More recently, Ref.~\cite{ccgt2} advocates peeling light fermions off the Planck brane in order to reduce
$S$. This amounts to shift the couplings of light fermions to the gauge bosons, reabsorbing in the process
some, or even all, of $S$. 
Finally, one might take the result in eqn.~(\ref{svsn}) as an indication that $N$ must be small. This 
pushes the theory into the realm of a strongly coupled KK sector. This is the scenario entertained in 
Ref.~\cite{bn}. The result are theories where the KK sector is not well defined since KK states are not
narrow, well spaced resonances. In this case, there is no gap between the TeV scale and the cutoff 
of the 5D $AdS_5$ theory where we could defined individual, weakly coupled states. We would expect one 
broad resonance encompassing all the KK states. Above the cutoff of a few TeV, stringy dynamics come 
into play. This scenario is quite reminiscent of a Walking Technicolor theory. This can be seen in the 
schematic phase diagram of Figure~\ref{phase}, where the 't Hooft coupling $g^2 N/16\pi^2$ is plotted
against energy.    
Here, $\Lambda$ is the energy scale where the CFT group exhibits non-trivial IR dynamics. 
In the large $N$ limit, it is possible to calculate $S$ reliably in the 5D $AdS_5$ theory. 
This is not the case in the Technicolor and Walking Technicolor theories. 
However, as $N$ is taken to be smaller, reliability may be lost in the 5D theory too. 
In any case, large or small $N$, the electroweak corrections come mostly from $E\sim \Lambda$, where
all theories will give similar results. However, at higher energies the theories may be quite different, 
with the dual of the $AdS_5$, a conformal theory with a 't Hooft coupling above 1 all to way to high 
energies. The question remains whether or not our knowledge of these differences can be put to use
to improve our understanding of strongly coupled theories at the TeV scale. 
Perhaps, EWSB is a consequence of a 4D conformal theory and the study of 5D theories could 
help illuminate some of its technical aspects.

%%%%%%%%%%%%%%%%%%%%%%%%%%%%%%%%%%%%%%%%%%%%%%%%
%% BACKMATTER,
%%%%%%%%%%%%%%%%%%%%%%%%%%%%%%%%%%%%%%%%%%%%%%%%

\begin{theacknowledgments}
I thank the organizers for a very interesting meeting and
for the opportunity they have given me to participate. 
\end{theacknowledgments}

%%%%%%%%%%%%%%%%%%%%%%%%%%%%%%%%%%%%%%%%%%%%%%%%
%% You may have to change the BibTeX style below, depending on your
%% setup or preferences.
%%
%% If the bibliography is produced without BibTeX comment out the
%% following lines and see the aipguide.pdf for further information.
%%
%% For The AIP proceedings layouts use either
%%%%%%%%%%%%%%%%%%%%%%%%%%%%%%%%%%%%%%%%%%%%

%\bibliographystyle{aipproc}   % if natbib is available
%\bibliographystyle{aipprocl} % if natbib is missing

%%%%%%%%%%%%%%%%%%%%%%%%%%%%%%%%%%%%%%%%%%%
%% You probably want to use your own bibtex database here
%%%%%%%%%%%%%%%%%%%%%%%%%%%%%%%%%%%%%%%%%%%
%\bibliography{sample}

%%%%%%%%%%%%%%%%%%%%%%%%%%%%%%%%%%%%%%%%%%%
%% Just a reminder that you may have to run bibtex
%% All of it up to \end{document} can be removed
%% if you don't like the warning.
%%%%%%%%%%%%%%%%%%%%%%%%%%%%%%%%%%%%%%%%%%%
%\IfFileExists{\jobname.bbl}{}
% {\typeout{}
%  \typeout{******************************************}
%  \typeout{** Please run "bibtex \jobname" to optain}
%  \typeout{** the bibliography and then re-run LaTeX}
%  \typeout{** twice to fix the references!}
%  \typeout{******************************************}
%  \typeout{}
% }

\end{document}